%% file: paper.tex
\documentclass[preprint]{jcgt}

\usepackage[dvipsnames]{xcolor}
\usepackage{multirow}
\usepackage{amsmath}
\usepackage{booktabs}
\usepackage{caption}
\usepackage{float}
\setciteauthor{Anas, Wolfe}
\setcitetitle{A Texture Lookup Approach to Bézier Curve Evaluation on the GPU}

\submitted{\today}

\accepted{yyyy-mm-dd}{yyyy-mm-dd}{yyyy-mm-dd}{Editor Name}{volume}{issue}{1}{6}{yyyy}
\seturl{http://jcgt.org/published/vol/issue/num/}

\begin{document}

\title{A Texture Lookup Approach to Bézier Curve Evaluation on the GPU}

\author{
\begin{minipage}[t]{0.48\textwidth}
\centering
Muhammad Anas~\href{https://orcid.org/0009-0007-7348-9934}{\includegraphics[width=8pt]{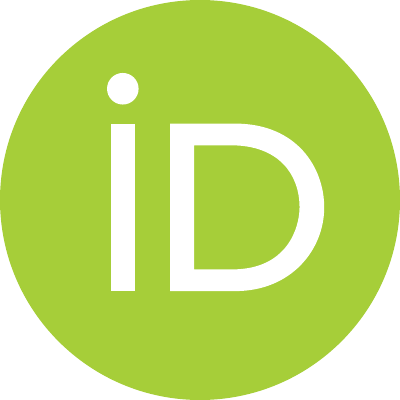}}\\
NUST (National University of Sciences and Technology), Pakistan
\end{minipage}
\hfill
\begin{minipage}[t]{0.48\textwidth}
\centering
Alan Wolfe~\href{https://orcid.org/0000-0001-9100-4928}{\includegraphics[width=8pt]{ORCIDlogo}}\\
SEED - Electronic Arts, USA
\end{minipage}
}

\teaser{
\begin{center}
\makebox[\textwidth][c]{%
\includegraphics[width=1\textwidth]{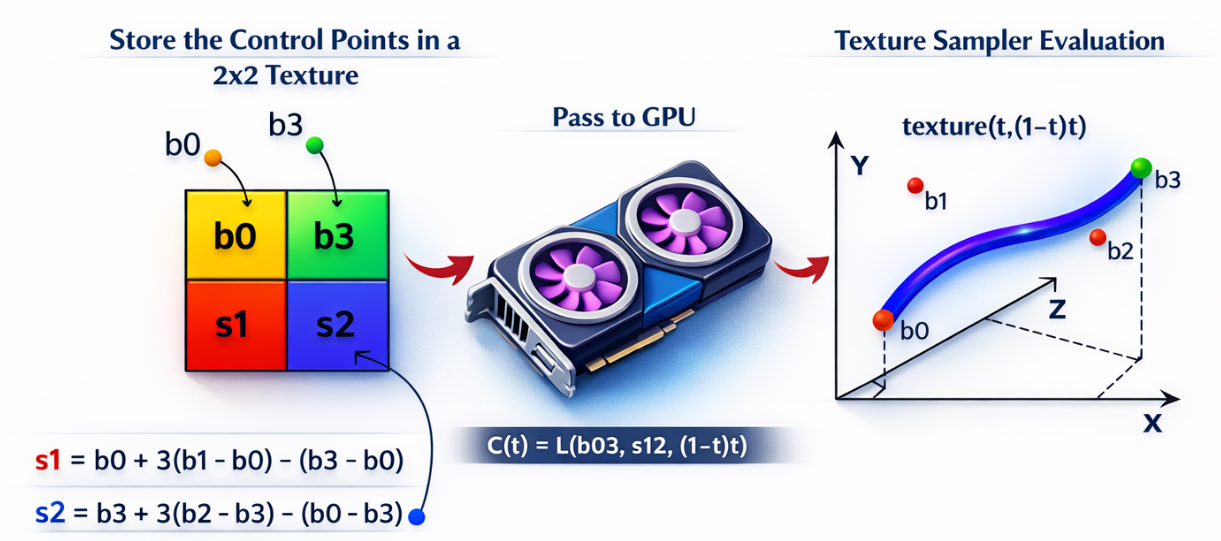}}
\end{center}
\caption{A 3D cubic Bézier curve is encoded in the RGB channels of a 2x2 texture and decoded using a single texture lookup with Seiler interpolation. The curve control points are
$b_0$, $b_1$, $b_2$, $b_3$, and the Seiler points are $s_1$ and $s_2$. This formulation extends to curves of higher and lower degrees, and allows computation to be offloaded to the
GPU's fixed-function texture interpolation hardware.}
\label{fig:teaser}
}

\maketitle
\thispagestyle{firstpagestyle}

\begin{abstract}
\small
We present a texture-based technique for evaluating Bézier curves on the GPU that leverages fixed-function linear texture interpolation hardware. By offloading curve
evaluation to the texture interpolator, this approach can improve performance in compute-bound GPU workloads. The method can also be used naturally for Bézier
surfaces and volumes and extends to advanced curve types such as B-splines, NURBS, and both integral and rational polynomials. We show how Seiler interpolation fits into
this framework to improve efficiency. We also compare performance and accuracy against curves evaluated as polynomials in shader code.
\end{abstract}

\section{Introduction}
\label{sec:introduction}

Bézier curves were invented in the late 1950s and early 1960s by Pierre Bézier and Paul de Casteljau to produce curves that were easily controlled and reproduced in computer-aided design
of automobile bodies. Since then, Bézier curves have been widely adopted in computer graphics for describing geometry and for modeling how data animates over time or in response to other
parameters. Bézier curves have also been extended to increase their expressiveness and flexibility, such as with B-splines and NURBS. 

Yuksel explained how Seiler interpolation can be used to evaluate polynomial curves \cite{10.1145/3641233.3664331}, including Bézier curves, and noted that GPU linear
interpolation could be used for the initial linear interpolation steps, with the remaining interpolations done in shader code.

In this paper, we present the details of evaluating Bézier curves using hardware-accelerated linear texture interpolation and show how to perform all interpolations on the hardware,
using both the de Casteljau algorithm and Seiler interpolation. We also compare performance and accuracy against curves evaluated as polynomials in shader code.

\section{De Casteljau Using Linear Texture Interpolation}

\input{Diagrams/BezDecasteljau/figure.tex}

The link between Bézier curves and linear texture interpolation lies in the de Casteljau algorithm, which is based on recursive linear interpolation. For a quadratic curve with control points
$A$, $B$, and $C$, the de Casteljau algorithm linearly interpolates between $A$ and $B$ to get point $AB$, between $B$ and $C$ to get point $BC$, and between $AB$ and $BC$ to get the final
point $ABC$, all with the same interpolation time $t$.

Similarly, we can make a 2x2 texture that has a row of pixels with values $A$, $B$, and a row with values $B$, $C$. When this texture is linearly sampled at $(t, t)$, there
are effectively two horizontal interpolations that produce $AB$ and $BC$, followed by a vertical interpolation between them that produces $ABC$. This equivalence is shown in Figure \ref{fig_tex_decasteljau}.

For a cubic curve, the de Casteljau algorithm linearly interpolates between two quadratic curve points $ABC$ and $BCD$ to make cubic curve point $ABCD$. If we make a 2x2x2 volume
texture that has a front slice encoding the quadratic Bézier curve $ABC$ and a back slice encoding $BCD$, sampling at $(t,t,t)$ gives point $ABCD$. For higher-order curves, further
interpolation levels can be done in the shader after the texture reads, since 4D and higher-dimensional textures are not supported on modern GPUs.

When linearly sampling textures, the pixel value is at the center of the pixel, so the $t$ values need to range from the center of one pixel to the next, instead of being in the usual
$[0,1]$ range. For a 2x2 texture, that means that $t$ is in the range $[0.25, 0.75]$.

\subsection{$C^0$-Continuous Piecewise Curves}

\input{Diagrams/ZigZag/figure.tex}

To store multiple quadratic curves in a single de Casteljau texture, we would naively need a 2x2 group of pixels for each curve. However, it can be observed that the diagonal texels
only need to average to $B$. If we have a 2x2 texture with rows $A, B$ and $C, D$, and we define $s = 1 - t$, bilinear interpolation can be written and expanded as:

\begin{equation}
\begin{split}
(As + Bt)s + (Cs + Dt)t \Rightarrow \\ As^2 + (B+C)st + Dt^2
\end{split}
\end{equation}

Doing the same for a texture with rows $A, E$ and $E, D$, we get:

\begin{equation}
As^2 + 2Est + Dt^2
\end{equation}

These are equal when $2E = B+C$ or equivalently when $E = (B+C)/2$.  Thus, when the diagonal values mismatch, the control point is defined as their average. If we use this fact
while also reading in a zig-zag pattern, we can add two pixels for each additional quadratic Bézier segment, instead of four, which reduces redundant memory storage. This is shown in Figure \ref{fig_ZigZag}.

Since a 2D texture can store arbitrary $C^0$-continuous quadratic Bézier curves, we can place two of them front-to-back into a 3D texture and interpolate to make a cubic curve. The pattern would
continue for higher order curves if higher dimensional textures were supported.

\subsection{Surfaces, Volumes, and NURBS}

Bézier tensor-product surfaces and volumes can be evaluated using this technique as well. If you are using the texture interpolator and the shader to evaluate an order-$O$ curve, you can split
that into order $M$ and $N$ such that $M+N=O$, to evaluate order $M$ with coordinate $t$ and order $N$ with coordinate $u$ to get the point on a surface.  The simplest incarnation of this
is sampling a 2D texture at coordinate $(t, u)$ to evaluate a bilinear Bézier surface. Similarly, for volumes, you can break the order $O$ into $M$, $N$ and $P$ such that $M+N+P=O$, to make
a Bézier volume. The pattern extends to hypervolumes as well.

If needing to evaluate multiple curves at once, you can use multiple color channels. A nice coincidence is that a bicubic surface parameterized by $(u,v)$ can fit into a 2x2x2 RGBA 3D texture.
Reading the 3D texture at $(u,u,u)$ evaluates a cubic Bézier curve in each color channel. The four resulting points are the control points of the cubic-curve isoline to be evaluated with $v$ in
the shader to get the final point on the surface.

With some preprocessing, it is also possible to evaluate NURBS. Rational splines can be achieved by dividing one integral spline point by another in the shader, and B-splines can be converted
to piecewise Bézier curves using Boehm's algorithm.

Lastly, as power basis polynomials can be converted to and from Bernstein basis (Bézier) polynomials, this technique can be used to evaluate general polynomials as well.

\subsection{Limitations}

\input{Diagrams/CurveAccuracy/figure.tex}

When using a low-bit-depth texture format, the values that control points can take are limited; for example, 8-bit textures are limited to 255 values, and to $[0,1]$ when using
a unorm format. Furthermore, the values read from the curves are also limited to the same range and discretization as the texture format! To deal with this limitation, it is possible to
rescale the curves to fit within the range. Alternatively, higher-bit-depth textures can be used, which can also help if 8-bit discretization of control points is too coarse.

Another source of discretization error is the texture sampling itself. When performing interpolation, the $(u,v)$ coordinates are commonly converted to a fixed-point X.8-bit
format, leading to unavoidable quantization errors \cite{CUDA_PG}. If this is a problem, doing some interpolation levels in the shader can help reduce the error, since shaders run at full
32-bit floating-point precision. The quantization errors are shown in Figure \ref{fig_curve_accuracy}.

\section{Seiler Interpolation}

Seiler interpolation provides an efficient method for evaluating polynomial curves using a reduced number of linear interpolations. 
For cubic Bézier curves, it requires only three linear interpolations, compared to the six required by the classical de Casteljau algorithm. 
This efficiency is achieved through the introduction of carefully constructed difference terms.

A purely interpolation-based formulation can be obtained by defining the Seiler points
\[
s_1 = b_0 + d_1, \qquad s_2 = b_3 + d_2.
\]

We then compute
\[
b_{03} = L(b_0, b_3, t), \qquad
s_{12} = L(s_1, s_2, t),
\]
and evaluate
\[
C(t) = L\!\left(b_{03}, s_{12}, (1-t)t\right).
\]

This form uses only three lerps.

\subsection{Arbitrary Degree Curves}

For a polynomial curve of degree $d$ with control points 
$b_0, b_1, \dots, b_d$, Seiler interpolation can be written recursively as

\[
C(t) = L(b_0, b_d, t) + (1 - t)t \, D_1(t),
\]

where the recursive term is defined as

\[
D_i(t) =
\begin{cases}
0, & \text{if } 2i = d + 1, \\[6pt]
d_i, & \text{if } 2i = d, \\[6pt]
L(d_i, d_{d-i}, t) + (1 - t)t \, D_{i+1}(t), & \text{otherwise}.
\end{cases}
\]

The difference terms are computed from the Bézier control points. 
For general degree $d$:

\[
d_1 = d(b_1 - b_0) - (b_d - b_0),
\]
\[
d_{d-1} = d(b_{d-1} - b_d) - (b_0 - b_d).
\]

For degrees up to $5$, additional terms can be computed as:

\[
d_2 =
\binom{d}{2}(b_2 - b_1)
- \binom{d-2}{2}(b_1 - b_0)
- (d-3)(b_{d-1} - b_d)
- 3(b_{d-1} - b_1),
\]

\[
d_{d-2} =
\binom{d}{2}(b_{d-2} - b_{d-1})
- \binom{d-2}{2}(b_{d-1} - b_d)
- (d-3)(b_1 - b_0)
- 3(b_1 - b_{d-1}).
\]

For a degree-$d$ polynomial, Seiler interpolation requires $d$ linear 
interpolations, $\lceil d/2 \rceil$ of which can be computed in parallel, 
since they operate on disjoint pairs of difference terms.

Explicit closed-form difference rules are currently known up to degree $5$.
\section{Seiler Interpolation Using Linear Texture Interpolation}

Using Seiler interpolation, a cubic Bézier curve can be written as follows:

\begin{equation}
C(t) = L(b_0, b_3, t) + L(d_1, d_2, t)\,st
\end{equation}

Here, $L(a,b,t)$ denotes linear interpolation from $a$ to $b$ at parameter $t$, and $s = 1-t$. Observing that $A+Bt$ (a multiply-add) can be reformulated as a linear interpolation $L(A, A+B, t)$ allows us to rewrite the equation using only lerps.

\begin{equation}
C(t) = L( L(b_0, b_3, t), L(d_1 + b_0, d_2 + b_3, t), st)
\end{equation}

This is equivalent to the "pure lerp" form in Yuksel \cite{10.1145/3641233.3664331}, where $d_1 + b_0$ and $d_2 + b_3$ are replaced with $s_1$ and $s_2$, respectively.

\begin{equation}
C(t) = L( L(b_0, b_3, t), L(s_1, s_2, t), st)
\end{equation}

This allows us to make a 2x2 texture where the top row of pixels contains $b_0$ and $b_3$, and the bottom row contains $s_1$ and $s_2$. Sampling that texture at coordinate $(t, st)$ gives the final point on the curve. This is shown in Figure \ref{fig:teaser}.

Using Seiler interpolation in this way allows us to store a cubic curve in a 2x2 texture, whereas the de Casteljau texture formulation would require a 2x2x2 texture. A cubic curve has 4 control points, and a 2x2 texture has the same number of degrees of freedom with 4 pixels, so
there is no redundant storage, unlike the de Casteljau formulation which requires 8 pixels.

The highest-order curve currently known in Seiler interpolation is a quintic curve, which has 6 control points. The pure-lerp form can be written as:

\begin{equation}
    \begin{split}
    C(t) = L(b_{05}, L(b_{05} + d_{14}, b_{05} + d_{14} + d_{23}, st), st) \\
    \text{where } \\
    b_{05} = L(b_0, b_5, t) \\
    d_{14} = L(d_1, d_4, t) \\
    d_{23} = L(d_2, d_3, t)
    \end{split}
\end{equation}

The second parameter of the outermost linear interpolation can be realized with a 2D texture that is read at location $(t, st)$:

\[
\begin{bmatrix}
b_0 + d_1 & b_5 + d_4 \\
b_0 + d_1 + d_2 & b_5 + d_4 + d_3
\end{bmatrix}
\]

Knowing that we must interpolate from $b_{05}$ to this value in the outermost lerp, we need a 2D texture that, when read at $(t, st)$, gives us $b_{05}$. We can do that with this texture:

\[
\begin{bmatrix}
b_0 & b_5\\
b_0 & b_5
\end{bmatrix}
\]

When this texture is the first slice of a volume texture, and the previous texture is the second slice, we can sample the volume texture at $(t, st, st)$ to get the point on the quintic curve at time $t$, using a single trilinear texture read.
See Table \ref{tab:SeilerLayout} for a diagram of Seiler textures of various degrees.

Just like in the de Casteljau formulation, when linearly sampling textures, the pixel value is at the center of the pixel. The interpolation-time values need to range from the center of one pixel to the next, instead of being in the usual
$[0,1]$ range. For a 2x2 texture, that means that $t$ is in the range $[0.25, 0.75]$. For the combined term $st$, the conversion happens after the multiplication.

\begin{table}[H]
\centering
\renewcommand{\arraystretch}{1.5}

\begin{tabular}{|c|c|c|}
\hline
\textbf{Degree} & \textbf{$z=0$ Layer} & \textbf{$z=1$ Layer} \\
\hline

Quadratic ($d=2$) &
$\begin{bmatrix}
B_0 & B_2 \\
B_0 + d_1 & B_2 + d_1
\end{bmatrix}$
&
N/A\\
\hline

Cubic ($d=3$) &
$\begin{bmatrix}
B_0 & B_3 \\
B_0 + d_1 & B_3 + d_2
\end{bmatrix}$
&
N/A\\
\hline


Quartic ($d=4$) &
$\begin{bmatrix}
B_0 & B_4 \\
B_0 & B_4
\end{bmatrix}$
&
$\begin{bmatrix}
B_0 + d_1 & B_4 + d_3 \\
B_0 + d_1 + d_2 & B_4 + d_3 + d_2
\end{bmatrix}$

\\
\hline

Quintic ($d=5$) &
$\begin{bmatrix}
B_0 & B_5 \\
B_0 & B_5
\end{bmatrix}$
&
$\begin{bmatrix}
B_0 + d_1 & B_5 + d_4 \\
B_0 + d_1 + d_2 & B_5 + d_4 + d_3
\end{bmatrix}$
\\
\hline

\end{tabular}

\caption{Structured arrangement of Seiler difference terms for polynomial degrees 2 through 5. The 2D textures are read using coordinates $(t, st)$, while the 3D textures are read using coordinates $(t, st, st)$. This layout allows efficient evaluation of curves using the GPU's texture interpolation hardware.}
\label{tab:SeilerLayout}
\end{table}

\section{Performance Analysis}

Modern GPUs hide memory latency by suspending the active thread and switching to another thread while waiting on memory reads. This allows memory and computation to happen in parallel, but GPU workloads are often unbalanced and end up either compute-bound or memory-bound. Our technique offloads
computational work from the shader program to the texture sampling unit, effectively trading compute for cache-friendly memory access of small textures. A necessary condition for our technique to provide a performance boost is that the workload must be compute-bound.
That condition isn't sufficient, though, because not all memory reads can be parallelized with all compute.

As a benchmark, we wrote a simple path tracer that traces $N$ rays per pixel per frame, has no textures, and uses hardcoded geometry to remove memory reads and force a compute-bound workload. After rendering, we use curves to color-grade the result. The first part of the work is fully compute, and the second part is entirely memory
reads when using our texture interpolation method. When profiling that workload, we found that performance decreased when using our texture interpolation method instead of compute, because this work was inherently serial and was difficult for the GPU to parallelize.

After modifying the path tracer to color-grade the result of each ray before accumulation, we saw a performance improvement because the next ray can start calculations while the previous ray is color grading. This work has more opportunity to parallelize memory and compute,
so the overall wall-clock time is reduced. These results are shown in Table \ref{tab:ColorGradePerf}.

\begin{table}[h]
\centering
\renewcommand{\arraystretch}{1.3}
\begin{minipage}[t]{0.38\linewidth}
\vspace{0pt}
\centering
\includegraphics[width=\linewidth]{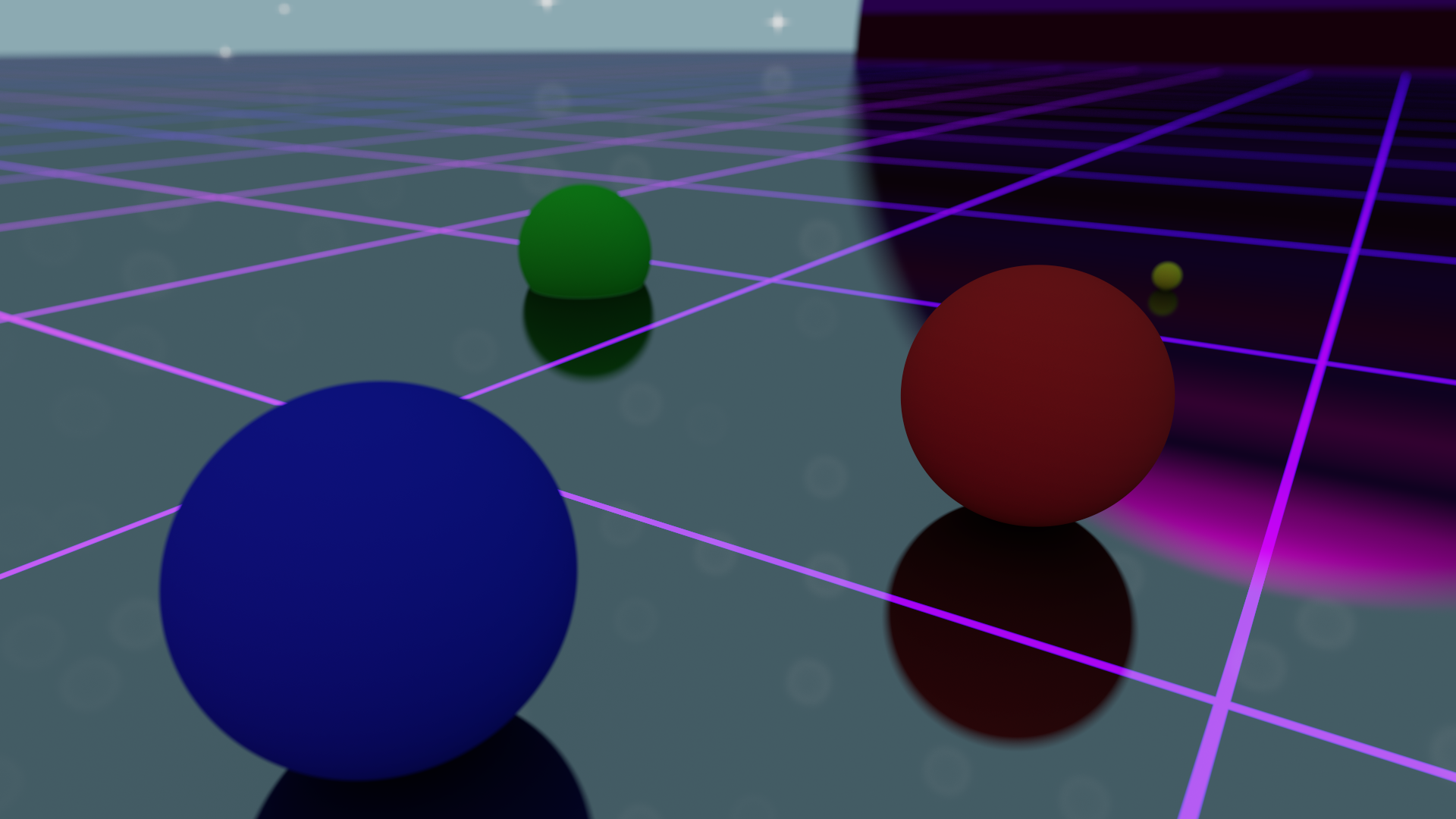}
\end{minipage}
\hfill
\begin{minipage}[t]{0.58\linewidth}
\vspace{0pt}
\centering
\begin{tabular}{lcc}
\hline
\textbf{Method} & \textbf{RTX 4090} & \textbf{RTX 3070} \\
\hline
Polynomial         & 1.952\,ms & 12.662\,ms \\
\textbf{Seiler Texture (Ours)} & \textbf{1.938\,ms} &  \textbf{12.496}\,ms \\
\hline
\end{tabular}
\end{minipage}
\caption{Frame times when path tracing a textureless scene of hardcoded geometry with 16 rays per pixel at 1920x1080 and using quintic curves for color grading each ray result. Averaged over 1000 frames. The RTX 3070 used stable power state to get consistent results.
Polynomial evaluates the curve as shader instructions, while Seiler Texture uses the GPU texture interpolator to evaluate the curve. Total wall-clock time is reduced when parallelizable work can be offloaded to the texture interpolator in compute-bound workloads.}
\label{tab:ColorGradePerf}
\end{table}

As another performance test, we implemented a particle simulation where several cubic Bézier curves are chained together to form a ring shape. Particles are spawned on the curves and move along them over their lifetime in a vertex shader.
We tested the performance using both our Seiler texture method, which stores each cubic curve in a 2x2 texture and evaluates the curve with a single bilinear read, and the Bernstein polynomial method, which evaluates
the curve as shader instructions. The results are shown in Table \ref{tab:seiler_comparison}.

\begin{table}[h]
\centering
\renewcommand{\arraystretch}{1.3}
\begin{minipage}[t]{0.38\linewidth}
\vspace{0pt}
\centering
\includegraphics[width=\linewidth]{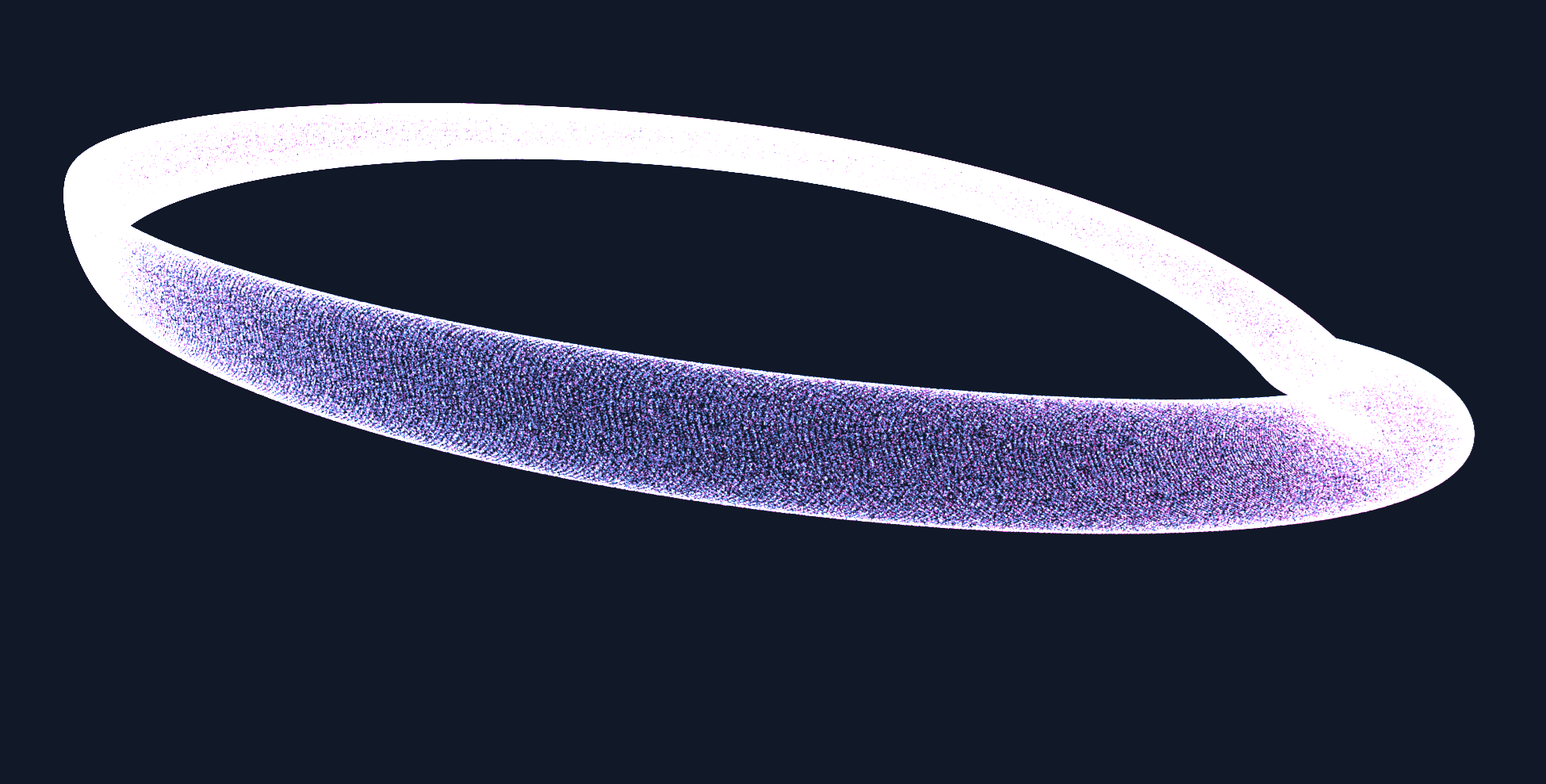}
\end{minipage}
\hfill
\begin{minipage}[t]{0.58\linewidth}
\vspace{0pt}
\centering
\begin{tabular}{lccc}
\hline
\textbf{Method} & \textbf{1m} & \textbf{5m} & \textbf{10m} \\
\hline
Polynomial         & 6.9\,ms & 13.9\,ms & 23.2\,ms \\
\textbf{Seiler Texture (Ours)} & 6.9\,ms &  \textbf{10.3}\,ms & \textbf{18.6}\,ms \\
\hline
\end{tabular}
\end{minipage}
\caption{Frame time for a particle simulation with particles moving along cubic Bézier curves, measured on an RTX 4050. Polynomial
evaluates the curve as shader instructions, Seiler Texture uses the GPU texture interpolator to evaluate the curve.}
\label{tab:seiler_comparison}
\end{table}

\section{Conclusions and Future Work}

We have shown that certain types of operations can be offloaded to the GPU's fixed-function texture interpolation hardware to improve performance in compute-bound GPU workloads.
While the quality of those calculations can degrade by being done at a low bit count quantized precision, the option to offload only some of the work allows a sliding scale of
performance benefits versus quality.

We show how to evaluate curves, surfaces, and volumes using the GPU texture interpolator to perform the de Casteljau algorithm, but we only show curves for Seiler interpolation.
Exploring surfaces and volumes with Seiler interpolation in the GPU texture interpolator could be fruitful.

Yuksel's paper provides formulas for Seiler interpolation up to degree 5, which has 6 control points. We believe that for a 3D volume texture with 8 pixels arranged as 2x2x2, degree
7 with 8 control points should be possible, much like cubic Bézier curves with 4 control points can be evaluated with a 2D texture with 4 pixels as 2x2. It may also be possible
to make higher-order curves in the same texture sizes by putting constraints on the control points and removing degrees of freedom, such as is done for piecewise curves that require
continuity at the endpoints. Exploring these avenues may lead to more use cases.

\subsection*{Acknowledgements}

We would like to thank the Rendering Engineering Special Interest Group (\#rendering-engineering-sig) chat room at Electronic Arts for helping us understand when these methods
would be useful.

\small
\bibliographystyle{jcgt}
\bibliography{paper}

\section*{Index of Supplemental Materials}

The Demos folder contains three subfolders.

\begin{itemize}
    \item \textbf{ColorGrading} - The path tracer used to benchmark the performance of color grading with curves, as shown in Table \ref{tab:ColorGradePerf}. Written in Gigi (https://github.com/electronicarts/gigi); also contains code-generated DX12 and WebGPU versions.
    \item \textbf{ParticleSim} - The particle simulation used to benchmark the performance of evaluating curves with particles moving along them, as shown in Table \ref{tab:seiler_comparison}.
    \item \textbf{ViewCurves} - The curve visualization utility shown in Figure \ref{fig_curve_accuracy}. Written in Gigi (https://github.com/electronicarts/gigi); also contains code-generated DX12 and WebGPU versions.
\end{itemize}

\section*{Author Contact Information}

\hspace{-2mm}\begin{tabular}{p{0.5\textwidth}p{0.5\textwidth}}
Muhammad Anas \newline
\href{mailto:hanas.bscs22seecs@seecs.edu.pk}{hanas.bscs22seecs@seecs.edu.pk}
&

Alan Wolfe \newline
\href{mailto:awolfe@ea.com}{awolfe@ea.com}

\end{tabular}

\afterdoc

\end{document}

%% file: Diagrams/BezDecasteljau/figure.tex
\begin{figure}
	\centering
	\includegraphics[width=\textwidth]{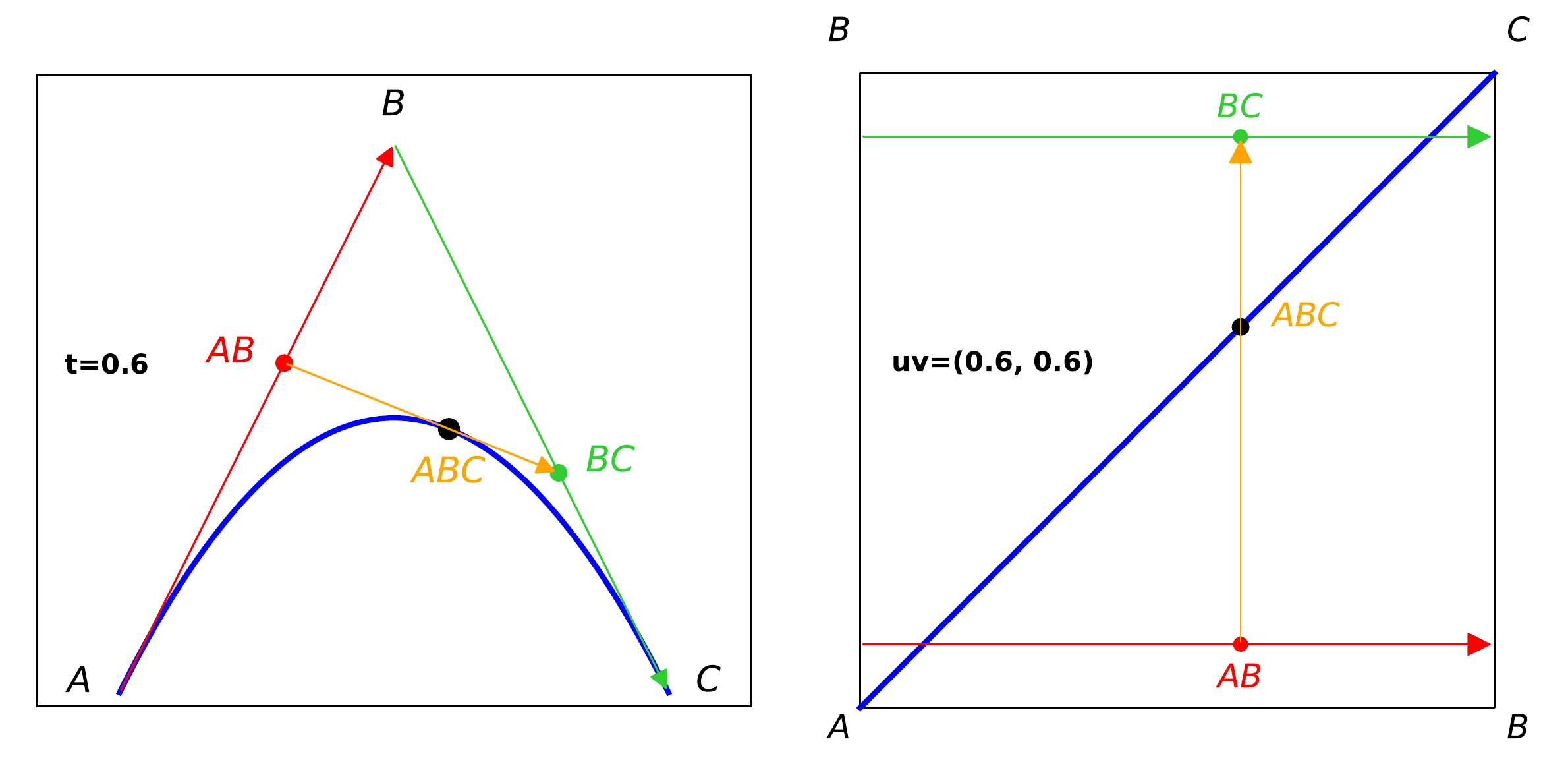}
	\caption{The de Casteljau algorithm for a quadratic Bezier curve with control points A, B and C (left). The equivalent bilinear interpolation setup (right).}
	\label{fig_tex_decasteljau}
\end{figure}

%% file: Diagrams/ZigZag/figure.tex
\begin{figure}
	\centering
	\includegraphics[width=\textwidth]{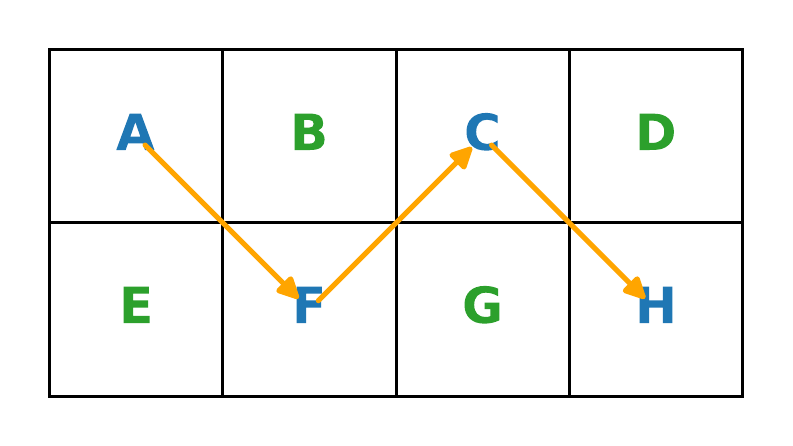}
	\caption{Three quadratic Bezier curves with C0 continuity stored in a 4x2 texture, instead of a 6x2 texture. The orange line shows how to sample the texture as t varies in $[0,3]$.
    The average of the green pixels opposite the orange lines define the middle control point. For $N$ curves, there will always be $N$ constraints with $N+1$ free variables to satisfy them.}
	\label{fig_ZigZag}
\end{figure}

%% file: Diagrams/CurveAccuracy/figure.tex
\begin{figure}
    \centering
    \includegraphics[width=\textwidth]{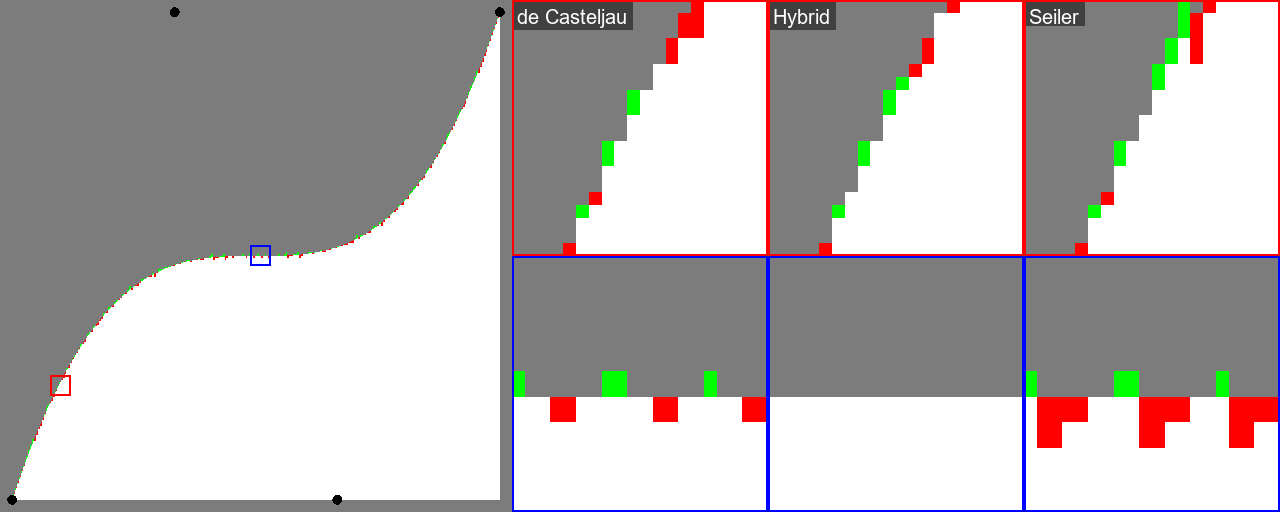}
    \caption{Using the texture interpolator to evaluate Bézier curves suffers from quantization errors due to fixed point quantization of the $(u,v)$ coordinates.
    Here a cubic curve is shown evaluated using the de Casteljau texture method which samples a 2x2x2 volume texture with a single trilinear texture read. A hybrid mode is shown
    which evaluates two quadratic curves with the texture interpolator (two bilinear reads) and then does the final lerp in the shader at full 32 bit floating point precision to reduce the error.
    Lastly the Seiler method is shown which takes a single bilinear sample of a 2x2 texture. In all images, the polynomial ground truth is shown in white. Where the curve being evaluated
    is less than the ground truth, the curve is shown in red, and where it is greater than the ground truth, it is shown in green. The full sized image shows the de Casteljau method compared to the ground truth.
    Seiler is most affected by quantization error, but uses the fewest texels.}
    \label{fig_curve_accuracy}
\end{figure}

%% file: paper.bib
@inproceedings{10.1145/3641233.3664331,
author = {Yuksel, Cem},
title = {Seiler's Interpolation for Evaluating Polynomial Curves},
year = {2024},
isbn = {9798400705151},
publisher = {Association for Computing Machinery},
address = {New York, NY, USA},
url = {https://doi.org/10.1145/3641233.3664331},
doi = {10.1145/3641233.3664331},
booktitle = {ACM SIGGRAPH 2024 Talks},
articleno = {60},
numpages = {2},
location = {Denver, CO, USA},
series = {SIGGRAPH '24}
}

@misc{CUDA_PG,
  author = {{NVIDIA Corporation}},
  title = {{NVIDIA CUDA C++ Programming Guide}},
  howpublished = {\url{https://docs.nvidia.com/cuda/cuda-c-programming-guide/\#linear-filtering}},
  year = {2025},
  note = {{Version 13.1 Update 1, as of January 2026}},
}
